\documentclass[a4paper,twoside]{article}

\usepackage{subfigure}
\usepackage{calc}
\usepackage{amssymb}
\usepackage{amstext} 
\usepackage{amsthm}
\usepackage[table,xcdraw]{xcolor}
\usepackage[fleqn]{amsmath} 
\usepackage{pslatex}
\usepackage{enumitem}
\usepackage{textcomp} 
\usepackage{url}
\usepackage{array,multirow,graphicx}

\usepackage{SCITEPRESS}    

\begin{document}

\title{Towards Readability Aspects of Probabilistic Mode Automata
}

\keywords{Software Engineering, Formal methods, Petri Nets}

 \author{ 
 \authorname{ 
  Heinz Schmidt, Maria Spichkova 
  }
  \affiliation{ 
  School of Science, RMIT University\\
  Melbourne, Australia\\
 $\{$heinz.schmidt, maria.spichkova$\}$@rmit.edu.au
  } 
 }

\abstract{
This paper presents a new approach and design
model targeting hybrid designer- and operator-defined performance
budgets for timing and energy consumption. The approach is based on Petri Nets formalism. 
As the cognitive load is typically high while using formal methods, this increases the chances of mistakes. 
Our approach is focused on the readability aspects and aims to decrease the cognitive load of developers. 
We illustrate the proposed approach on example
of a sample embedded multi-media system, a modern digital
camera.\footnote{%
Preprint. Accepted to 14th International Conference on Evaluation of Novel Approaches to Software Engineering (ENASE 2019). Final version published by SciTePress.
}   
}

\onecolumn 
\maketitle 
\normalsize \vfill

\newcommand\nat[1]{[0...#1-1]}
\newcommand\Nat{\mathbb{N}}
\newcommand\Real{\mathbb{R}}
\newcommand\Int{\mathbb{Z}}
\newcommand\Rat{\mathbb{Q}}
\newcommand\Trac{\mathbb{T}}
\newcommand\Power{\cal{P}}

%cpo related definitions
\newcommand{\weakereq}{\trianglelefteq}
\newcommand{\weaker}{\vartriangleleft}
\newcommand{\strongereq}{\trianglerighteq}
\newcommand{\stronger}{\vartriangleright}

\newcommand{\newterm}[1]{\textbf{#1}}
\newcommand{\tbf}[1]{\textbf{#1}}

\newcommand{\deff}{\/\ensuremath{=_{def}}\/}

\newcommand{\RFSM}[1][]{{\ensuremath{\text{CR-FSM}_{#1}}\/}}
\newcommand{\SRFSM}[1][]{{\ensuremath{\text{MR-FSM}_{#1}}\/}}
\newcommand{\MFSM}[1][]{{\ensuremath{\text{M-FSM}_{#1}}\/}}
\newcommand{\PFSM}[1][]{{\ensuremath{\text{P-FSM}_{#1}}\/}}
\newcommand{\CFSM}[1][]{{\ensuremath{\text{C-FSM}_{#1}}\/}}

\newcommand\set[1]{\ensuremath{\{{#1}\}}}

\newcommand\ipnote[1]{\par {\bf IP:}{\em #1}\par}
\newcommand\hwsnote[1]{\par {\bf HWS:}{\em #1}\par}
\newcommand\msnote[1]{\par {\bf MS:}{\em #1}\par} 
\newcommand{\Idle}{\emph{IDLE}}
\newcommand{\SF}{\emph{SF}}
\newcommand{\HS}{\emph{HS}}
\newcommand{\LS}{\emph{LS}}
\newcommand{\FE}{\emph{FE}}
\newcommand{\wcet}{\textit{WCET}}
\newcommand{\bcet}{\textit{BCET}}
\newcommand{\acet}{\textit{ACET}}
\newcommand{\wcec}{\textit{WCEC}}
\newcommand{\bcec}{\textit{BCEC}}
\newcommand{\acec}{\textit{ACEC}}
\newcommand{\Exposure}{\textit{Exposure}}
\newcommand{\Deployment}{\textit{Deployment}}
\newcommand{\Throttle}{\textit{Throttle}}
\newcommand{\Compr}{\textit{Compr}}
\newcommand{\ImgSz}{\textit{ImgSz}}
\newcommand{\BufSz}{\textit{BufSz}}
\newcommand{\auto}{\textit{AUTO}}
\newcommand{\CardSz}{\textit{CardSz}}
\newcommand{\MaxFrames}{\textit{MaxFrames}}
\newcommand{\frameWCET}{\textit{frameWCET}}
\newcommand{\EndWCET}{\textit{endWCET}}
\newcommand{\bufferCapacity}{\textit{bufferCapacity}}
\newcommand\doAS{\ensuremath{\mathrm{AS}}}
\newcommand\doIB{\ensuremath{\mathrm{IB}}}
\newcommand\doIS{\ensuremath{\mathrm{IS}}}
\newcommand\doIP{\ensuremath{\mathrm{IP}}}
\newcommand\refAUTO{\ensuremath{\mathrm{AUTO}}}
\newcommand\noShoot{\ensuremath{\mathrm{Shoot}}}
\newcommand\onShoot{\ensuremath{\mathrm{Shoot}}}
\newcommand\nxtBF{\ensuremath{\mathrm{nxt BF}}}
\newcommand\onBF{\ensuremath{\mathrm{on BF}}}
\newcommand\noBF{\ensuremath{\mathrm{no BF}}}
\newcommand\Motors{\ensuremath{\mathrm{Motors}}}
\newcommand\HSIMGPROC{\ensuremath{\mathrm{HS\_IMGPROC}}}
\newcommand\HSSHOOT{\ensuremath{\mathrm{HS\_SHOOT}}}
\newcommand\ShootSync{\ensuremath{\mathrm{Shoot\_Sync}}}
\newcommand\BFSync{\ensuremath{\mathrm{BF\_Sync}}}

%===============================================
\section{\uppercase{Introduction}}
\label{sec:introduction}

In the domain of embedded systems, the trend to enhance more and more system functionalities through software solution is constantly increasing. This makes the design of these systems and the corresponding quality assurance more and more challenging~\cite{Sangiovanni-VincentelliM01}. 
Real-time and dependability constrains provide additional challenges, which also lead to necessity of probabilistic analysis  
within the phases of design and
  verification of these  systems.
  Also, some constraints within embedded systems are mutually dependent, for example,  timing and energy consumption constraints cannot be analysed independently of each other, see \cite{mudge01power,saxe:power:cacm:2010,wolf:mpsocs:07}.

  One of the successfully applied  paradigms is 
  Component-based software development, which was initially introduced many decades ago~(CBSD, see \cite{Adler95,Clements95}). 
  However, CBSD  cannot solve directly  issues related to the constraints on safety, timing, energy consumption, etc.~\cite{DBLP:conf/fm/HenzingerS06,spichkova2012verified}, but can provide a solid basis for extended approaches.

  In recent work \cite{peake11qosa} 
  we have extended our rich
  architecture definition language (RADL, see \cite{Schmidt03Trustworthy})
  and underlying theory~\cite{SchmidtPXTKFB03} 
  to meet such industrial
  requirements, aiming at a scalable and compositional (component-based) approach
  to soft dependability guarantees: with probability, guarantee
  risk, execution time, cost etc. 
  Industrial practice
  requires the capability to compose a variety of heterogeneous models
  and components, specified and designed using different methods and 
  frameworks. Many real-world engineering environments are not locked
  into a single model, single framework or single-language
  environment. While we abstract from the programming languages
  underlying such an heterogeneous software engineering approach, we
  hope to show that, and how, our design-oriented model-based approach
  links with concrete programming by means of elementary modelling
  blocks providing abstractions directly for code blocks. This is
  natural and perhaps more appropriate in design of embedded systems
  than in other fields, as component models in this context often use
  architectural elements to abstract from software and hardware blocks
  at the same time. However we expect that this approach carries
  across to other domains.   
  
  In our current work, we  targeting hybrid designer- and operator-defined performance
budgets for timing and energy consumption. We propose an approach that is on Petri Nets formalism. 
Our approach is focused on the readability aspects and aims to decrease the cognitive load of developers, as having high cognitive load increases the chances of mistakes in system design and quality assurance process. 
We also aim to keep the method lightweight, following the classification presented in \cite{zamanskytowards}.

To illustrate the proposed approach, we use an example of a sample embedded multi-media system, a modern digital camera.  
This allows us to demonstrate how the time (and the
  ensuing synchronisation) and energy constraints can be analysed taking into account their mutual dependencies. We propose that
  extra-functional properties have to be considered from early
  performance requirement specification through to model-based testing
  and run-time verification.  Beside the compositional approach to
  reasoning about and testing such properties in a hybrid modelling
  environment, our contribution is in the separation of concern of
  different aspects of modelling and in context-dependent methods of
  reasoning about such properties.  Notably we have developed methods
  which allow automated contextual resource allocation strategies,
  under dynamically varying, and suitably parameterised, architectural
  configurations.

%===============================================
\section{\uppercase{Example: Digital Camera}}
\label{sec:example}

Consider the design of a modern digital camera
from the perspective of different types of use:

{\em Scenario 1:} A busy professional sports photographer requires the ability to capture many hundreds or thousands of high quality images rapidly, with minimal shutter lag, in rapid bursts of up to 100 photos.
Within the given price point afforded by budget, she is prepared to sacrifice ``convenience'' features, accepting shorter battery life and fewer shots per memory card while carrying extra battery packs, memory cards or even a laptop for frequent uploads, as well as extra lenses, and manage reconfiguration as needed.

{\em Scenario 2:} One weekend a family member is getting married, and as the {\em de facto} camera expert she has agreed to act as a semi-official or backup photographer for the wedding.
In this capacity she aims for simplicity and convenience,
so she can still enjoy the day and mingle without being conspicuous or weighed down by equipment.
The couple insist they prefer photos in a standard compressed consumer format (JPEG),
which at least eliminates extra effort later at her workstation, and maximises memory card capacity.
She selects what she can carry easily---a single camera body and lens and perhaps a single additional memory card, but no extra battery pack.
She is unwilling to spend anything like her usual time and effort on camera configuration, instead often (perhaps not always) relying on camera to automatically select exposure, focus and aperture.
Occasionally, for particularly important shots she takes full control again.
In this second case, battery life is paramount.

The specific challenge is to design a camera which is capable of flexible reconfiguration to suit multiple contexts, including for example these.
The generic challenge is to:
\begin{itemize}
    \item[(i)] {\em Characterise context} in terms of user configuration choices, usage (e.g. selected modes/operations/functions) and user-visible desired properties.
    \item[(ii)] {\em Reason in a context-sensitive way about system properties} and manage internal configuration to ensure consistency between configuration/profiles and desired properties.
For example, to make the camera battery last longer, the camera must somehow sacrifice quality and/or performance in an acceptable way. 
\end{itemize}
However the true usage context is often hard to predict.
What exactly are the user's requirements and intentions?
Even the user may not know exactly what she intends beyond the immediate moment.
Contextual uncertainty extends to environmental conditions, which may have a non-trivial impact on performance. For example ambient temperature may affect performance (e.g., energy consumption) of key camera components significantly, including batteries~\cite{Rao03}, sensors and actuators such as lenses.
This has implications for the design not only of embedded systems, but also at a macroscopic level.
Thus, large-scale computing centres have significant inter-dependency on their local environment;
such facilities are already planned with environmental conditions such as temperature in mind to be able to maximise performance and performance per cost while minimising cooling and energy consumption.

We extend the camera design presented by Lee~\cite{lee:B:2006:arch}.  
In our example, the camera has the following logical components: 
\begin{itemize}
    \item a general purpose processor (GPP), 
    \item a digital signal processor (DSP), 
    \item actuators to control, e.g., mirror and shutter curtain, lens focus and aperture, 
    \item sensors, e.g., for auto-focus, 
    \item a buffer to store images temporarily, and 
    \item a flash memory as a long-term storage media. 
\end{itemize}
To keep the example small enough for a conference paper,  we abstract from other typical functions such as USB driver for photo download, LCD user interface,
camera flashbulb, and various advanced settings.  

In high performance scenarios a dedicated GPP-flash memory link is possible.
We focus on the interplay of functionality relevant for taking a range of different shots involving real-time physical control, as well as selecting tradeoffs between timing and energy consumption.

\begin{figure}[htb]
\begin{center}
    \includegraphics[scale=0.65]{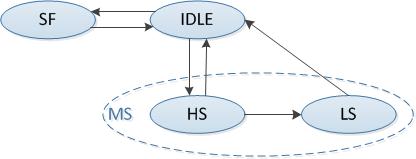}
%}
\end{center}
\caption{Digital Camera: Modes}
\label{fig:modes}
\end{figure}

As presented in Figure~\ref{fig:modes}, the system has three modes, each with different resource requirements:
\begin{itemize}
\item 
\Idle\ mode covers waiting for shutter half/full press and pre-focusing.
\item
In {\em single frame (SF)} mode, the camera returns to the
idle mode after shooting is completed, while 
\item
in {\em multi frame (MF)} mode, shooting is
continued as long as the shutter release button is kept pressed.
\end{itemize}
$MF$ contains two sub-modes, \emph{high-speed (HS)} and \emph{low-speed (LS)}.
$MF$ starts with $HS$ and switches to  $LS$ if/when the image buffer gets full,
where shooting of the consequent frame is delayed until enough space is
freed in the buffer by writing to the flash memory. 
With these mode abstractions in mind, from a design perspective it is expected
that refinements to components used in these modes may enable new features (for example
smart/continuous save in HS at a performance penalty).

Furthermore, in each mode the user can select lens focusing and exposure metering to be performed
automatically or manually, i.e. each mode has four submodes.  
More precisely, in the case of multi frame shooting, each of the $MF$ submodes, $HS$ and $LS$, has four further submodes:  
\begin{itemize}
\item \FE:  
automatic operations are  fully enabled: 
both autofocus $AF$ and automatic exposure  $AE$ are enabled; 
\item \emph{F:}  
only the autofocus $AF$ operation is enabled;
\item \emph{E:} 
only the automatic exposure  $AE$ operation is enabled;
\item 0:  
neither autofocus $AF$ nor automatic exposure  $AE$ are enabled.
\end{itemize}
In the \Idle\ mode the user may perform $AF$, $AE$ or both, while composing
a picture.
During this time DSP cannot be activated and $AF$ and $AE$
operations are performed on GPP to reduce energy consumption. When the
user presses the shutter release button, first, $AF$ and $AE$ operations
that are being executed are completed, then the idle mode is
terminated and the system switches to $SF$ or $MF$ depending on the user
selection.

\begin{figure}[htb]
\begin{center}
\includegraphics[scale=0.47]{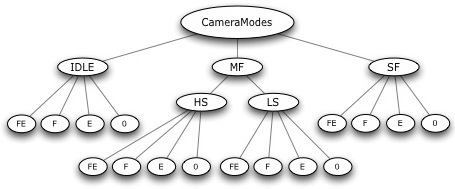}
\caption{Digital Camera: Submodes Hierarchy}
\label{fig:modesHierarhy1}
\end{center}
\end{figure}

Another way to represent system modes (which can be related to the same submodes hierarchy as introduced in Figure~\ref{fig:modesHierarhy1}) 
is to work parallel with on mode variables, because the choice to activate $AF$ and $AE$ operations is highly independent 
of whether the camera is in the \Idle, \SF or one of the multi frame modes. 
Let call them 
\emph{CameraMode} and \emph{AutoMode} defined over enumeration types
\[
  \{ \Idle,~ \SF,~  \HS,~  \LS \}
 \]
 and
\[\{
 \FE,~   F ,~   E ,~  0\}
 \]
 respectively. 
We can also see this as a feature composition/interaction, see e.g.,~\cite{Calder2000FeatureInt,Apel2010,Broy2010MSS}.
Thus, one feature is responsible for the choice of the current value of \emph{CameraMode}  
and for the processes in the corresponding mode, 
where the second feature solely deals with the $AF$ and $AE$ operations.

\begin{figure}[htb]
\begin{center}
\includegraphics[scale=0.55]{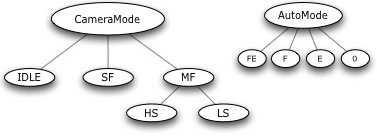}
\caption{Digital Camera: Parallel Model for the Submodes Hierarchy}
\label{fig:modesHierarhy2}
\end{center}
\end{figure}

Table \ref{table:components} lists some of the relevant software components, their descriptions and their implementation platform (GPP/DSP). Some components are implemented in both processors to allow dynamic re-configuration of the system in order to provide optimal resource usage. 
Within these constraints, a key challenge is allocating computing resources for the software elements to best suit partly predictable usage conditions.
The DSP is especially suited to image processing operations, yet the DSP has significant energy overheads.
We characterise the main design problems for the camera as follows.
(i) Given an overall objective (e.g. minimise time consumption), satisfy that objective at run time.
(ii) Given a usage profile, minimise energy and time consumption at run time.
 
\begin{table}[htb]
\begin{center}
{\footnotesize
\begin{tabular}{|l|l|c|c|} \hline
  & Description                            &GPP        &DSP       \\ \hline
\multicolumn{4}{|c|}{\bf Operatoions}\\\hline  
AF     & AutoFocus:                  &\checkmark &\checkmark \\
     &  Automatic lens  focusing                   &  &  \\
    \hline
  AE      & Automatic Exposure          metering          &\checkmark &\checkmark \\       
 \hline
 IP     &   Image Processing on     local                    &-          &\checkmark \\       
         &  buffer,  red-eye reduction, etc.      &           & \\ \hline
 IB     & Image     Buffering:                          &-          & \checkmark          \\ 
     &  Transfer image                        &         &           \\ 
         &  from  sensor to local buffer                 &           &           \\ \hline
 IS     & Image     Storage:                   &\checkmark &-          \\ 
   & Transfer images                  &  &          \\ 
         &  from  buffer to flash card                 &           &           \\ \hline
 AS     &   Activate Shutter etc.                    &\checkmark &\checkmark \\ 
     &     (e.g. aperture adjust)                 &  &  \\ 
 \hline
 BC     & Buffer check:                 &-          &\checkmark \\  
      & Check if buffer is full                  &           &  \\ \hline
\multicolumn{4}{|c|}{\bf Modes}\\\hline
IDLE    &  Idle mode &- &\checkmark \\\hline
 SF     &  Single-Frame shutter &\checkmark&\checkmark \\\hline
 MF     & Multi-Frame  shutter &\checkmark&\checkmark \\\hline
\multicolumn{4}{|c|}{\bf Submodes}\\\hline 
 FE     &  AF and AE enabled &\checkmark &\checkmark \\\hline
 F     &  AF only enabled &\checkmark &\checkmark \\\hline
 E     &   AE only enabled &\checkmark &\checkmark \\\hline
 0     &   AF\&AE disabled &\checkmark &\checkmark \\\hline
\end{tabular}
}
\end{center}
\caption{Software Components}
\label{table:components}
\end{table}

%===============================================
\section{\uppercase{Proposed visualisation approach}}
\label{sec:representation}

\begin{figure}[htb]
    \includegraphics[width=0.48\textwidth]{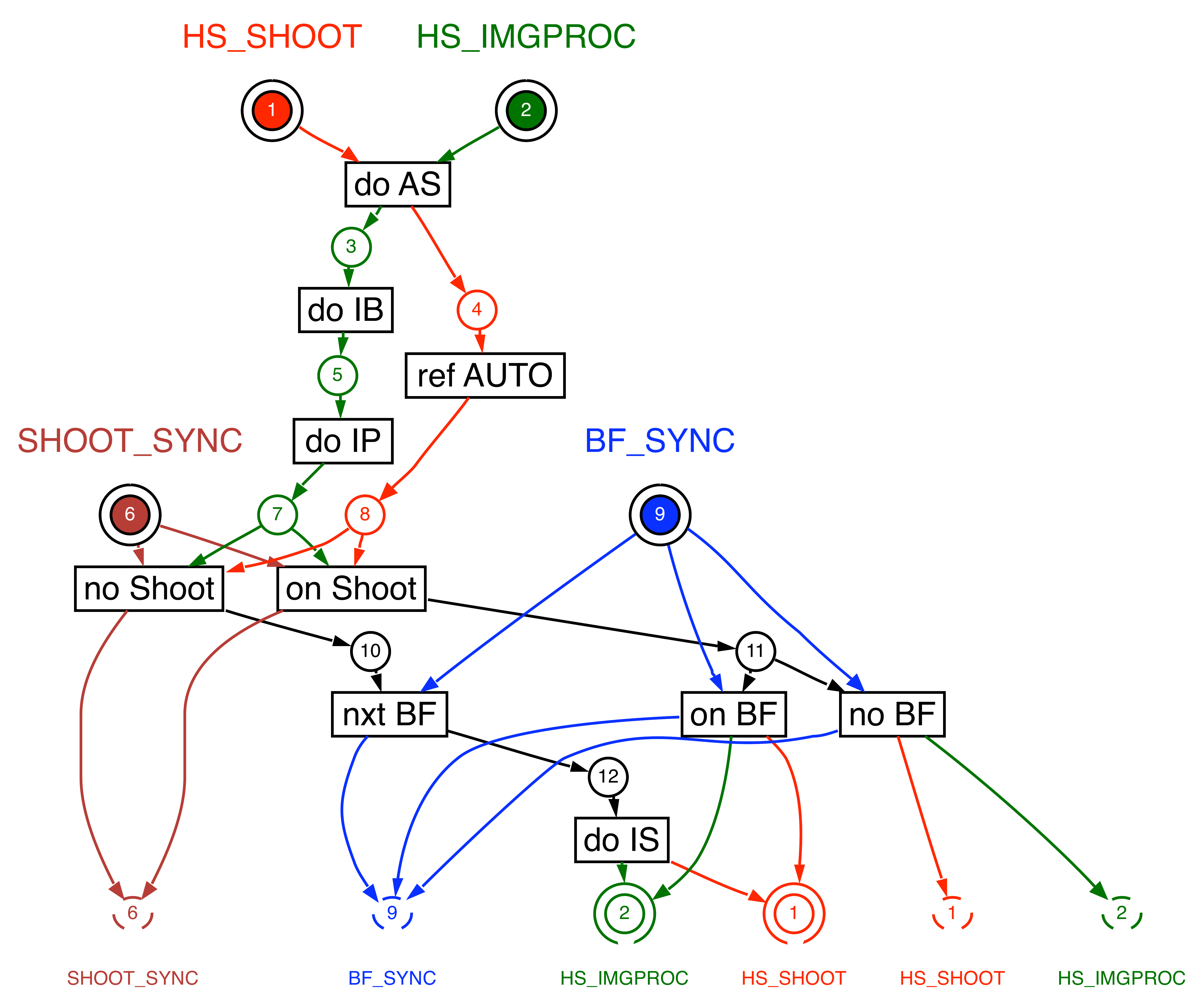}
\caption{$HS$ mode details presented as a coloured Petri net}
\label{fig:HS}
\end{figure}

One of the problems using formal representation is that often only two factors are considered as important: 
the method must be sound and give such a representation, which is concise and
beautiful from the mathematical point of view, without taking into account any question of
readability, usability, etc., but even small syntactical changes of a method can make it more understandable and usable for  
engineers \cite{Constantine2003,Dhillon2004,Klare2000,Spichkova2013HFFM}. 
 Figure~\ref{fig:HS} presents an the example of Petri net specifying HS mode details for the digital camera, 
which provides a typical representation of a coloured Petri net. 
Within our approach, we propose the following enhancements: 
To make representation more readable, first of all we should take into account the human factor. 
Thus, if a path (in this case a colour marked path, green or red) starts on the left/right of the net, 
we should proceed to draw it on the same side if possible
and avoid cross moving the paths without any important reason. 

\begin{figure}[htb]
\begin{center}
\includegraphics[scale=0.65]{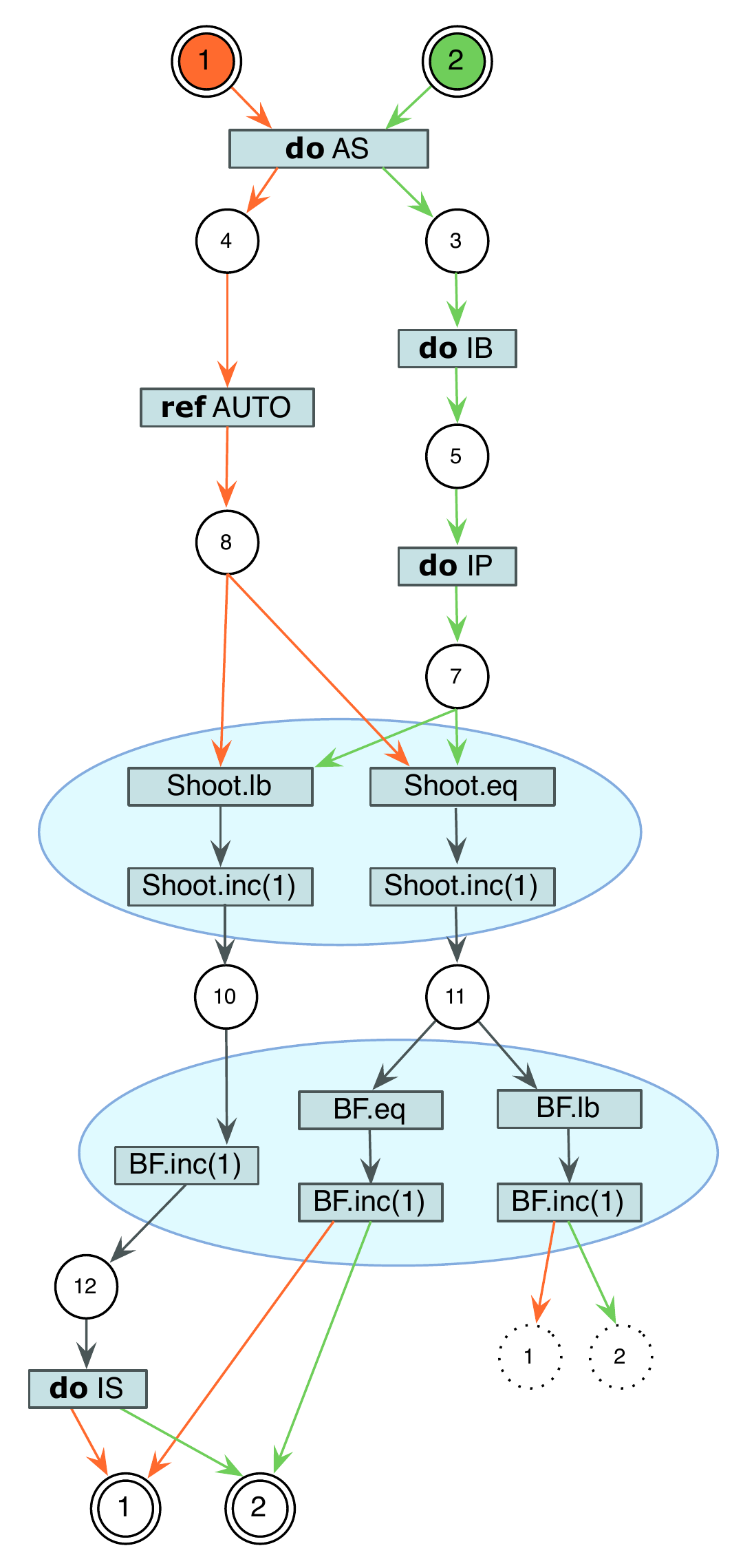}
\caption{HS mode details: Another Visualisation}
\label{fig:HSmodesVis1}
\end{center}
\end{figure}

Thus, on Figure~\ref{fig:HS} two paths are switched after the operation \emph{do AS}, which can confuse some readers.
Then, we can try to find a solution to avoid a lot of crossing arrows having different meanings: 
the blue and maroon arrows indicate synchronisation of the counters, 
and we can replace them by visual grouping of operations on the same counter. 
As result we obtain an optimised coloured Petri net presented in Figure~\ref{fig:HSmodesVis1}, 
which is semantically equivalent to the representation in Figure~\ref{fig:HS}.
This optimisation increases ease of use by human readers (designers, testers etc.) 
without decreasing simplicity for machine readability and semi-automated support or
 expressiveness/power (for the domain or domains of choice).

Usability derives from the following aspects:
\begin{itemize}
\item 
Lowering the barrier between the simplified and expressive language
for the machine support and that of the domain languages of the
user(s) and associated with the purpose, e.g., by using  
controlled natural languages that try to avoid disadvantages of both natural and formal languages 
and being a subset of a natural language with a well-defined syntax and semantics, see 
 \cite{kuhn2010doctoralthesis,Fuchs_specifyinglogic,PulmanNaturalLanguageProcessing}. 
\item
Applying an appropriate automatisation of a number of steps within the modelling and verification process: this not only saves human time and allows to get results much faster then humans can produce manually,
 but also (partially) excludes  the human element as the most ``unreliable” in failure, see~\cite{Redmill1996,spichkova2016human}. 
 For example, a formal specification can be generated from the corresponding CASE tool representation which can be 
 edited in a more readable   
 way also using predefined templates, see~\cite{spichkova2013we,vo2016model}.
\item 
Supporting directly common and standard abstractions that are
well-established (and hence part of the software engineering training), 
e.g. \emph{Message Sequence Charts}~\cite{Blaise2008,MSC:Kluwer2003}), 
or defined in standards (such as UML, IEC-61131, etc):
\item
Unification of the representation of any information we are dealing with (see, e.g., \cite{Spichkova2013HFFM});
\item
Easing the use of novel compositional principles and high-level
tools, that are opening novel and powerful methods to users of formal
specification or specification-based/model-driven methods.
\end{itemize}

Having a representation like presented on Figure~\ref{fig:HSmodesVis1}, 
we can easily transform a Petri net to a \emph{hierarchical  MSC}.
In the case a component-based specification of the system is need in addition to the process representation, 
an MSC can be schematically translated to the corresponding formal specification as shown 
in~\cite{spichkova2012verified,dentum_focus_tb}. 
Let us also shortly discuss translation/representation of the following modelling artefacts: 
(global) parameters, local, time and counter variables.

\emph{Local variable} use  can be translated into state and transition
label expansions for  NF purposes~\cite{LTS2005}, 
but can also be
intuitively understood in data types and data structures that capture
state. % 
However for Petri net
normal form used in our approach and compositionality considerations restrictions need to
be designed along with such capabilities to limit the scopes of these
variables appropriately, viz. to  FSM components of nets, in terms of
their use in guards and assignments associated with transitions and
states.

\emph{Global parameter} use are of a similar nature with respect to
normalisation but needs to be limited to achieve compositionally. 
For example, we could say the global parameters may occur locally in guards (i.e., they are read-only)
as well as in initialisation expressions 
(for the initial states when FSM objects are created) or with re-assignments limited to
higher-level FSMs (such as mode automata) when submodes are entered
and before these branch out into rational parallel processes.
Another common example is the use of iterator and bounded loop
process constructs that have a very structured use of local \emph{counter}
variables which never serve synchronisation but are providing a
reasoning tool for local termination and performance approximation,
based on an interplay of local (loop) invariants and loop control
variables, which implies strict monotonicity and boundedness.

\emph{Time variables (clocks)} are a further example and in some sense a
special case of counter control above -- in the sense that all
practical approaches to timed automata and synchronous time models
discretise an infinite number of real-time points into a real-time
intervals with integer bounds and then solve a linear convex hull
problem to determine feasibility and/or optimal schedules that meet
time constraints. 
There is also a significant difference here, that
needs to be considered, relative to counters. 
In general, counter
processes can be explained as a macro structure based on sequence and
choice, and hence are lower-level automata (or process expressions)
themselves, and hence they do not add 'new' semantics but can be
explained in terms of existing semantics.
For example if we are in rational parallel processes, they are just a
syntactic sugar extension that does not take us out of this class.
Likewise with other classes of processes (such as pushdown automata).
In contrast, timed extension are true semantic extensions, in that
they define a different class of behaviours and automata, because the
define \emph{what} the legitimate processes (occurrence nets) are that are
traces of the give language (net system or process expression).

%===============================================
\section{\uppercase{Related Work}}
\label{sec:related}

Component-based software engineering utilises a well-defined
composition theory  to enable the prediction of such properties.
as performance and reliability.
This is one of the largest fields of software and system engineering, 
there are many approaches on component-based design (CBD) covering different aspects and focusing on requirements, quality, timing properties etc.  
(see e.g., \cite{CBSQ2003,Broy2010MSS,Broy99alogical,cocome}). 
Several  component-based prediction approaches, 
e.g.\ Palladio~\cite{Kapova2010,Martens2008,Palladio2007}, 
CB-SPE~\cite{CB-SPE}, 
ROBOCOP~\cite{Robocop}
 (see also a survey in~\cite{Becker06performanceprediction})  derive the benefits of 
 reusing well-documented component specifications. 
 In our approach we focus on the questions of resource-awareness and 
 adaptivity of systems as well as on the readability aspects of the formalism.

Mode automata have a long history motivated by real-time design
practices and methods used in industry in connection with statecharts.
Maraninchi et al.~\cite{maraninchi03mode} capture
the notion of modes formally for a practical extension of the
real-time synchronous language Lustre and include elements of the
well-known I/O-automata.  Mode automata define synchronous mode
automata as a hybrid between data-flow and transition systems. 
Talpin
et al.~\cite{talpin06polychronous} extend this work to so-called
polychronous mode automata to work with the multi-clock data-flow
formalism SIGNAL.  Both these types of automata are non-deterministic
and do not deal with probabilities.  The (bisimulation) equivalence
and therefore compositional reasoning for mode automata is
undecidable.  However, Maranichi et al.\ introduce a synchronised
(lock-step) parallel product for modes in which shared symbols
(intersection of alphabets nonempty) are synchronised while local
symbols (the symmetric difference of the alphabets) are independent.
While the modes of a single automaton are mutually exclusive in their
approach, and the behaviour of these mode automata is fully abstract
wrt.\  probabilistic testing, the automata product suffers from
combinatorial explosion (state space explosion), due to the aim of
allowing arbitrary shared variables and interference of parallel
processes.

Cheung et al.~\cite{cheung12} describe an architecture-level
method, SHARP, for predicting reliability (and timing) of concurrent
systems.  Whereas SHARP is specifically designed for reliability and
timing prediction, our method is intended to be generic thus also
catering, e.g., for energy consumption.  SHARP models involve {\em
  scenarios} which are either {\em basic} (similar to message sequence
charts) or hierarchical, involving sequential, conditional or
concurrent composition.  SHARP supports concurrent composition of
finite numbers of instances of a particular scenario, corresponding to
symmetrically replicated components.  SHARP derives completion time
and reliability predictions from scenarios for use at higher levels of
abstraction.  For each basic scenario, SHARP requires transition rates
for all individual actions, then calculates a single continuous-time
Markov model model from which completion time and reliability are
derived.  For an hierarchical scenario a system level CTMC is
constructed using abstraction techniques such as queuing networks and
abstraction of sequential components into single global states. 
In contrast, our approach requires probabilities/rates at the system
level only.  Our approach seeks to avoid or defer calculation of
monolithic models.  

Our cost estimation is inspired by Valiant's bulk synchronous-parallel
model~\cite{Valiant1990} of parallel computing where global {\em
  strong synchronisation} conservatively approximates systems which
may in reality use more fine grained synchronisation and indeed may
allow for more asynchrony than the above approximation would suggest.
In performance benchmarks reported in~\cite{Yusuf2011}, Yusuf et al. 
demonstrated that such conservative predictions may still be accurate enough
if there is enough WCET variation and a large enough number of
activities/tasks scheduled on individual processing elements.  Thus
adjacent modes may be assumed to be strongly separated in the global
model while in fact such modes are partially interleaved with respect
each other (subject to restrictions on repetition such as boundedness
for message sequence graphs as described by Alur~\cite{AlurY99} and
star-connectivity in trace languages).  For conservative cost
estimation purposes this seems reasonable.
We expect that (with diminishing returns) such models can be refined selectively,
to bound costs of adjacent sequences of overlapping modes, in a context-dependent way.

An interesting approach on integration of synchronous and asynchronous communication 
was presented by Hennicker  et al.~\cite{Hennicker2008, Hennicker2011}. 
In this approach, 
 I/O-transition systems  were used as the formal background for modelling of system behaviour. 
As result,  a refinement relation was defined, which is compositional w.r.t. synchronous and asynchronous connections
of components and which preserves connection-safety, and next
existing interface theories for modal I/O-transition systems were extended to support assemblies, (greybox)
assembly refinement and assembly encapsulation, also showing  that communication-safety is preserved
by assembly refinement, that black-box refinement of component interfaces
is compositional w.r.t.\ grey-box refinement of assemblies and, conversely, that assembly
encapsulation maps grey-box to black-box refinement.

%===============================================
\section{\uppercase{Conclusions}}
\label{sec:conclusions}

In this paper, we proposed a Petri-Nets-based approach targeting hybrid designer- and operator-defined performance
budgets for timing and energy consumption. 
The core focus of this approach is on decreasing the cognitive load of the designers to decrease the chances of design mistakes. 
To achieve better readability, we extended the coloured Petri Nets formalism. 
To illustrate the proposed solution, we presented an example
of a sample embedded multi-media system, a modern digital
camera.  

\emph{Future work:} We are going to integrate the presented approach with the results of our prior work, a probabilistic global behaviour analysis approach
 developed for reliability and fault-tolerance studies (including fault
 injection) and a parallelism/concurrency focused framework centred on
 partially ordered traces, Petri nets and timing/energy costs.

\bibliographystyle{abbrv}
 
%\bibliography{biblio.bib} 

\end{document}